\documentclass[aps,prl,twocolumn]{revtex4}

\usepackage{epsfig}
\usepackage{amssymb}
\usepackage{graphicx}

\begin{document}
\title{Time-dependent Andreev reflection}
\author{Dmitri V. Averin$^{1}$, Gongqi Wang$^{1}$, and Andrey S. Vasenko$^{2,3}$}
%\email{dmitri.averin@stonybrook.edu}
\affiliation{$^{1}$Department of Physics and Astronomy, Stony Brook University, SUNY,
Stony Brook, NY 11794-3800 \\
$^{2}$National Research University Higher School of Economics, 101000 Moscow, Russia \\
$^{3}$I.E. Tamm Department of Theoretical Physics, P.N. Lebedev Physical Institute, Russian Academy of Sciences, 119991 Moscow, Russia}

\begin{abstract}

We extend the basic theory of Andreev reflection (AR) in a normal metal/superconductor junction
to the situation with an arbitrary time-dependent bias voltage $V(t)$ across the junction. The central element of the theory is the fact that the Fourier transform of the AR amplitude has a casual structure. As an example, the theory is used to describe the current response to short pulses of the bias voltage, which create coherent superposition of quasiparticle states with different energies. The current oscillates in time, with the gap frequency $\Delta/\hbar$, and also as a function of the pulse area $\int V(t)dt$, with the period of the single-electron flux quantum $e/h$.
\end{abstract}

\maketitle

Andreev reflection (AR) \cite{Andreev1964} is the process of conversion of electrons in a
normal metal (N) into Cooper pairs in a superconductor (S) and vice versa, and represent the
main mechanism of electron transport across an NS interface with large electron transparency.
As a result, AR determines the basic transport characteristics of the NS junctions, including
the linear conductance \cite{Kastalsky1991,Wees1992}, average current \cite{Blonder1982,Volkov1993,Hekking1993} and current noise \cite{Muzykantskii1994,Galaktionov2009}; in junctions with very low transparency, one can observe individual AR transitions \cite{Maisi2011}. AR also gives rise to an enormous amount of various other transport phenomena. To give just a few examples, it is the basic mechanism of the supercurrent flow in Josephson junctions \cite{Kulik1978,Furusaki1991,Beenakker1991} and, in the form of multiple Andreev reflections (MAR), determines all their transport characteristics at finite bias voltages: average
current \cite{Klapwijk1982,Arnold1987,Gunsenheimer1994,Bratus1995,Averin1995,Cuevas1996,Scheer2001}, current noise \cite{Averin1996,Averin1999,Jehl1999,Ronen2016} and full statistics of charge transfer \cite{Cuevas2003,Johansson2003}. AR produces thermoelectric effects in NS junctions \cite{Bardas1995,Brinkman2003,Rajauria2008,Vasenko2010,Pekola2014}, and plays an important role in superconducting structures with other materials, e.g., carbon nanotubes \cite{Buitelaar2003,Jarillo-Herrero2006,Hata2018}, graphene \cite{Beenakker2008,Du2008,Mizuno2013,Efetov2016}, topological insulators \cite{Badiane2011,Finck2014,Jauregui2018,Jonckheere2019}.
In all these situations, AR is typically considered under the conditions of the constant bias voltage, when the energy of the quasiparticles which determine the AR amplitude can be taken to be constant throughout the scattering process. However, in many situations, a more detailed theory of AR with an arbitrary time-dependent bias voltage $V(t)$ is desirable. The primary goal of this work is to develop such a theory.

Physically, the main novel feature of the time-dependent bias is the creation of coherent quantum superposition of the quasiparticle states with different energies. As a result, the junction produces  an oscillatory current response, the magnitude of which is sensitive to the coherence properties of these superpositions. For instance, as shown below, the oscillating current generated by the AR processes is more stable to thermal averaging than that produced by the quasiparticle tunneling, since AR always involve two quasiparticles with vanishing total energy. Also, in the limit of short pulses of the bias voltage, the quantum superposition of quasiparticle energies leads to the current oscillations as a function of the total ``magnetic flux'' $\int V(t)dt$ carried by the pulse
with the period of the single-electron flux quantum $h/e$.

We begin by outlining the derivation of our main results. We use the most basic model of an NS junction, a small constriction between the normal and superconducting electrodes (Fig.\ 1a)
with all transport modes characterized by one transparency $D$. For short constriction, it is
possible to consider the constriction region itself as normal, reducing the transport inside
it to the motion of independent quasiparticles. We assume that the bias voltage $V(t)$ across the constriction is arbitrary, but varies on the time scale set by the energy gap $\Delta$ of the superconductor and other small energies in the problem, e.g., temperature $T$; i.e., its characteristic frequencies are smaller than the microscopic energy scales set by the Fermi
energy and by the traversal time of the barrier that determines the transparency $D$.
In this regime, we can neglect the effect of the bias voltage on $D$, and also use the
quasiclassical approximation for the quaiparticle motion through the constriction.
This general approach is similar to the one used to describe the time-dependent transport in
normal conductors -- see, e.g., \cite{Moskalets2012}.

\begin{figure}[t]
\centering
\includegraphics[width=0.4\textwidth]{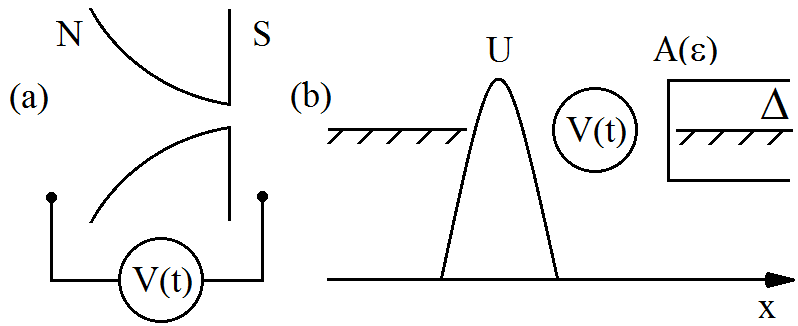}
\caption{{\protect\footnotesize {Sketch of (a) a short constriction between a
normal metal (N) and a superconductor (S) used as the basic model of an NS junction,
and (b) a quasiparticle scattering scheme in the junction that consists
of the normal scattering with the scattering matrix $U$ ($U^*$ for the holes), energy
change of the incident quasiparticles according to  Eq.~(\ref{acc2}) due to the
time-dependent bias voltage $V(t)$, and Andreev reflection at the NS interface with
the amplitude $A(\epsilon)$ (\ref{AR}). Also indicated schematically in (b) is the
Fermi level in the two electrodes. } }}
\label{fig1} \end{figure}

One starts by accounting for the effect of the bias voltage $V(t)$ on the quasiparticle motion
between the two electrodes. The voltage creates the electric field $E(t,x)$ localized in the constriction, $V(t)=\int dx E(t,x)$, where $x$ is the coordinate along the constriction. (The assumption of the relatively low frequencies of $V(t)$ made above also implies that $E(t,x)$ is  quasistatic from the point of view of electromagnetism.) Describing this field through the vector potential $A(t,x)$, $E=-\partial A/\partial t$, and solving the time-dependent Scr\"{o}dinger
equation with $A(t,x)$ in the quasiclassical approximation, we see that the amplitude $\psi_e (t)$ of the wavefunction of an electron crossing the constriction from the normal metal to the superconductor acquires the $A(t,x)$-dependent phase:
\begin{eqnarray}
\psi_e (t) \rightarrow \exp \{ \frac{ie}{\hbar} \int dx A(t,x)\} \psi_e (t) \nonumber \\
= \exp \{ \frac{-ie}{\hbar} \int^{t} dt' V(t')\}\psi_e (t)  = e^{-i\phi(t)} \psi_e (t)\, ,
\label{acc1} \end{eqnarray}
where the phase $\phi$ is defined by the relation $\dot{\phi} = \frac{e}{\hbar} V(t)$.
Electrons passing through the constriction in the opposite direction accumulate the phase
of the opposite sign. The phases acquired by the holes are switched in comparison to those
for the electrons. Physically, these phase factors describe the acceleration/deceleration
of the quasiparticles as they move between the junction electrodes. In the quasiclassical approximation, the particle energies change, while the change of their velocity is small
and is neglected.

Next, we introduce the Fourier components of the accumulated phase:
\begin{equation}
e^{-i\phi(t)} =\int d\omega W(\omega)e^{-i\omega t},
\label{ff} \end{equation}
in close analogy to what is done in the ``Werthamer theory'' of the time-dependent properties
of the Josephson tunnel junctions \cite{Werthamer1966,Larkin1966}. Then, the electron acceleration
process (\ref{acc1}) has the following form in terms of the energy components $a(\epsilon)$ of the wavefunction $\psi_e (t)$:
\begin{equation}
a(\epsilon) \rightarrow \int d\omega W(\omega)a(\epsilon-\omega)\, .
\label{acc2} \end{equation}

The quasiparticle acceleration/deceleration described above should be combined with the standard quasiparticle scattering scheme in the NS junction as illustrated schematically in Fig.\ 1b. For electrons incident from the normal electrode on the constriction at energy $\epsilon$, the scattering process consists of the normal barrier scattering characterized by the scattering matrix $U$,
\begin{equation}
U= \left( \!\begin{array}{cc}  r \, ,& t \\  t' \, ,& r'\end{array}\! \right)\, , \;\;\;\; r'=-t'r^*/t^*,
\label{scat} \end{equation}
with $|t|^2=|t'|^2=D$ and $|r|^2=|r'|^2=R$, and Andreev reflection at the NS interface with the amplitude:
\begin{equation}
A(\epsilon) =\left\{ \!\!\begin{array}{ll}  x-\mbox{sgn}(x)\sqrt{x^2-1}  , & |x|>1\, ,
\\  x-i\sqrt{1-x^2} ,&  |x|<1\, ,  \end{array} \right. \;\;\;\; x=\epsilon/\Delta.
\label{AR} \end{equation}
For the holes incident from the normal electrode, the AR amplitude is the same, while the scattering matrix is $U^*$. Solving this scattering scheme for the electrons and the holes, taking the standard average over their equilibrium energy distributions at temperature $T$, and combining the electron and the hole contributions to the current, we get the total current in the constriction:
\begin{eqnarray}
I(t) = \frac{eN}{2 \pi} \int d\omega d\nu W^*(\omega)W(\nu)e^{i(\omega-\nu)t}
\int d\epsilon [f(\epsilon-\omega)\nonumber \\ -f(\epsilon+\nu)]
\cdot \Big[1+D^2\frac{A^*(\epsilon)}{1-[A^*(\epsilon)]^2R} \frac{A(\epsilon+\nu-\omega)}{1-A^2(\epsilon+\nu-\omega)R} \nonumber \\
-R\frac{1-[A^*(\epsilon)]^2}{1-[A^*(\epsilon)]^2R} \frac{1-A^2(\epsilon+\nu-\omega)}{1-A^2(\epsilon+\nu-\omega)R}\Big] , \;\;
\label{curr} \end{eqnarray}
where $N$ is the number of the spin-degenerate transport modes in the constriction.
In principle, this expression can be used directly to calculate the current in the NS junction.
For instance, for constant bias voltage, $W(\omega)=\delta(\omega-eV/\hbar)$, and Eq.~(\ref{curr}) reduces to the well-known expression \cite{Blonder1982},
%
%\begin{eqnarray}
%I = \frac{eN}{2 \pi \hbar}  \int d\epsilon [f(\epsilon-eV)-f(\epsilon+eV)]
% \nonumber \\ \cdot \Big[1+D^2\frac{|A(\epsilon)|^2}{|1-A^2(\epsilon)R|^2}
%-R\Big|\frac{1-A^2(\epsilon)}{1-A^2(\epsilon)R}\Big|^2 \Big] , \;\;
%\label{curr1} \end{eqnarray}
%
which describes the evolution of the dc $IV$ curves from the regime of the tunnel to the
ballistic junction with increasing quasiparticle transparency $D$. For general time-dependent
voltage, however, it is more convenient to transform Eq.~(\ref{curr}) explicitly into the
time domain.

As the first step in this direction, we separate the term in Eq.~(\ref{curr}) that does not
decay at large $|\epsilon|$. The magnitude of this term in the integral over $\epsilon$ is
$1-R=D$ and it gives $D\int d\epsilon [f(\epsilon-\omega)-f(\epsilon+\nu)]=D (\omega+\nu)$, meaning that this contribution in Eq.~(\ref{curr}) corresponds to the normal-state current $I_N(t)$ in the junction:
\begin{eqnarray}
I_N(t) = \frac{eN D}{2 \pi} \int d\omega d\nu (\omega+\nu) W^*(\omega)W(\nu)
e^{i(\omega-\nu)t} \nonumber \\   =\frac{eN D}{\pi} \Big(-i\frac{\partial}{\partial t}e^{i\phi(t)}\Big)e^{-i\phi(t)}=GV(t)\, .
\label{norm} \end{eqnarray}
Here $G=e^2ND/\pi \hbar$ is the normal-state conductance.

After the separation of the normal-state part, the current can be expressed as
\begin{eqnarray}
I =I_N+ \frac{eN}{2 \pi} \int d\epsilon f(\epsilon)\Big[D^2\Big| \int d\omega e^{-i\omega t}
\frac{W(\omega) A(\epsilon+\omega)}{1-A^2(\epsilon+\omega)R}\Big|^2 \nonumber \\
+RD^2 \Big| \int d\omega e^{-i\omega t} \frac{W(\omega) A^2(\epsilon+\omega)}{ 1-A^2(\epsilon +\omega)R}\Big|^2 +2DR \nonumber \\\cdot \mbox{Re} \Big\{ e^{i\phi(t)} \int d\omega
e^{-i\omega t}\frac{W(\omega)A^2(\epsilon+\omega)}{1-A^2(\epsilon +\omega)R}\Big|^2 \Big\}
- ... \Big] , \;\;
\label{curr2} \end{eqnarray}
where the ellipsis denotes the subtracted identical terms in which $A(\epsilon+\omega)$
is replaced everywhere with $A^*(\epsilon-\omega)$. In this expression, we can transform the
AR amplitude into the time domain. More precisely, we introduce the two response functions
that enter Eq.~(\ref{curr2}):
\begin{eqnarray}
K(\tau) =\frac{i}{2\pi} \int dx e^{-ix\tau} \frac{A(x)}{1-A^2(x)R} \, , \label{corr1} \\
L(\tau) = \frac{1}{2\pi} \int dx  e^{-ix\tau} \frac{A^2(x)}{1-A^2(x)R} \, , \label{corr2}
\end{eqnarray}
where $\tau$ is the time normalized to the gap frequency, $\tau=t\Delta/\hbar$, and the
prefactors are chosen for later convenience.

An important property of the AR amplitude $A(x)$ (\ref{AR}) is that it can be viewed as the
reduction to the real axis of the function of the complex variable $z$: $A(z)=z-\sqrt{z^2-1}$.
The function $A(z)$ is analytic on the whole $z$ plane except for the cut on the $[-1,1]$
interval of the real axis, and $A(x)$ (\ref{AR}) on this interval is the value of $A(z)$ on
the upper (Im$z>0$) branch of the cut. This property implies that the AR amplitude in the time
domain has a clear casual structure, vanishing for $\tau<0$:
\begin{eqnarray}
\frac{i}{2\pi} \int dx e^{-ix\tau} A(x)=\Theta (\tau)\frac{i}{2\pi} \oint_C dz e^{-iz\tau} A(z)
\nonumber \\ =\Theta (\tau)\frac{2}{\pi} \int_0^1 dx \sqrt{1-x^2} \cos x\tau  = \Theta (\tau) \frac{J_1(\tau)}{\tau}\, , \label{ARF} \end{eqnarray}
where $C$ is the contour going clock-wise around the branch cut, and $J_1$ is the Bessel
function. One can see directly that the denominator in the functions (\ref{corr1}) and
(\ref{corr2}) does not add any poles to these functions, and therefore, they have the same
analytical properties as the AR amplitude $A(z)$. From this, we obtain directly the following expressions:
\begin{eqnarray}
K(\tau) =\Theta (\tau) \frac{2(1+R)}{\pi} \int_0^1 dx \frac{\sqrt{1-x^2}
\cos x\tau }{(1+R)^2-4x^2R} \, , \label{corr3} \\
L(\tau) = -\Theta (\tau) \frac{4}{\pi} \int_0^1 dx \frac{x\sqrt{1-x^2}
\sin x\tau}{(1+R)^2-4x^2R} \nonumber \\ =\frac{2}{(1+R)}
\frac{\partial K(\tau)}{\partial \tau} \, , \;\;\; \tau>0\, .\; \label{corr4}
\end{eqnarray}

Taking the inverse Fourier transform to express the amplitudes $W(\omega)$ in terms
of $\phi (t)$, and using the functions (\ref{corr3}) and (\ref{corr4}) in Eq.~(\ref{curr2}),
we obtain our main general result for the current in the NS junction driven by an
arbitrary time-dependent bias voltage, expressed directly in the time domain:
\[ I(t) = I_N+\frac{eNT\Delta}{\hbar^2} \Big\{ \frac{D^2\Delta}{\hbar}
\int_{-\infty}^{t} dt'dt'' \frac{\sin[\phi(t')-\phi(t'')]}{ \sinh[\pi T(t'-t'')/\hbar]}
\]
\vspace*{-2ex}
\begin{eqnarray} \cdot [K(t-t')K(t-t'')-R L(t-t')L(t-t'')] \nonumber \\
+2DR \int_{-\infty}^{t} dt' \frac{\sin[\phi(t)-\phi(t')]}{\sinh[\pi T(t-t')/\hbar]}
L(t-t') \Big\} . \;\;
\label{td} \end{eqnarray}

Equation (\ref{td}) for the time-dependent NS current can be used to calculate the current
under many different bias conditions. For intance, it shows that the main qualitative
feature of the junction response to the voltage that varies rapidly on the time scale
$\hbar /\Delta$ is interference of the quasiparticle reflection from the gap edges which
leads to oscillations of the current in time with the gap frequency, and also to oscillations
in magnitude with the applied voltage. Consider the simplest model of the voltage pulse that
is infinitely short on the gap time scale:
\[ V(t) =\Phi \delta (t)\, ,\;\;\;\; \phi(t) =2\pi \phi_V \Theta (t) \, ,\]
where $\Phi$ is the total area under the voltage pulse, which has the meaning of the
magnetic flux carried by this pulse, and $\phi_V\equiv e\Phi/h$ is the magnitude of this
flux in units of the single-electron flux quantum. Equations (\ref{norm}) and (\ref{td})
show that the current in this case is
\begin{equation}
I(t) = G \Phi \delta (t)+\Theta (t) I_0(t) \sin (2\pi \phi_V) \, ,
\label{delta} \end{equation}
where qualitatively, the current $I_0(t)$ oscillates and decays on the time scale $\hbar
/\Delta$. The magnitude of this oscillatory response is modulated periodically, with the
period of the single-electron flux quantum, by the area $\Phi$ of the bias voltage pulse.

Quantitatively, the current $I_0(t)$ is
\begin{eqnarray}
I_0(t) = \frac{2eNT\Delta}{\hbar^2} \Big\{ \frac{D^2\Delta}{\hbar}
\int_{0}^{t} dt'\int_{-\infty}^{0} \frac{dt''}{ \sinh[\pi T(t'-t'')/\hbar]}
\nonumber \\ \cdot [K(t-t')K(t-t'')-R L(t-t')L(t-t'')] \nonumber \\
+DR \int_{-\infty}^{0} dt' \frac{L(t-t')}{\sinh[\pi T(t-t')/\hbar]}
\Big\} . \;\;
\label{ampl} \end{eqnarray}
Equation (\ref{ampl}) can be evaluated explicitly at large temperatures, $T\gg \Delta$. In this limit,
different terms in (\ref{ampl}) have two different behaviors as functions of temperature. The
last term, which corresponds to the quasiparticle tunneling in the tunnel-junction limit, decays exponentially with $T$:
\begin{equation}
I_0^{(qp)}(t) = \frac{4G\Delta}{e}RL(t)e^{-\pi Tt/\hbar}.
\label{TLT} \end{equation}
This temperature dependence reflects the fact that the quasiparticles tunnel at different energies, and thermal averaging of the partial amplitudes at different energies results in the exponential decay. In contrast to this, the other two terms in Eq.~(\ref{ampl}) decay only as $1/T$, as a result of all AR processes having total zero energy, and therefore, thermal averaging reducing only the probability of the incident quasiparticle to have the initial energy within the range of $\Delta$. Indeed, in the limit $T\gg \Delta$, the first two terms in Eq.~(\ref{ampl}) give at times $t$
larger than $\hbar/T$, when the exponentially decaying quasiparticle contribution (\ref{TLT})
can be neglected:
\begin{equation}
I_0 (t) = \frac{\pi G\Delta^2D}{2eT}[K^2(t)-R L^2(t)].
\label{TLAR} \end{equation}

\begin{figure}[t]
\centering
\includegraphics[width=0.45\textwidth]{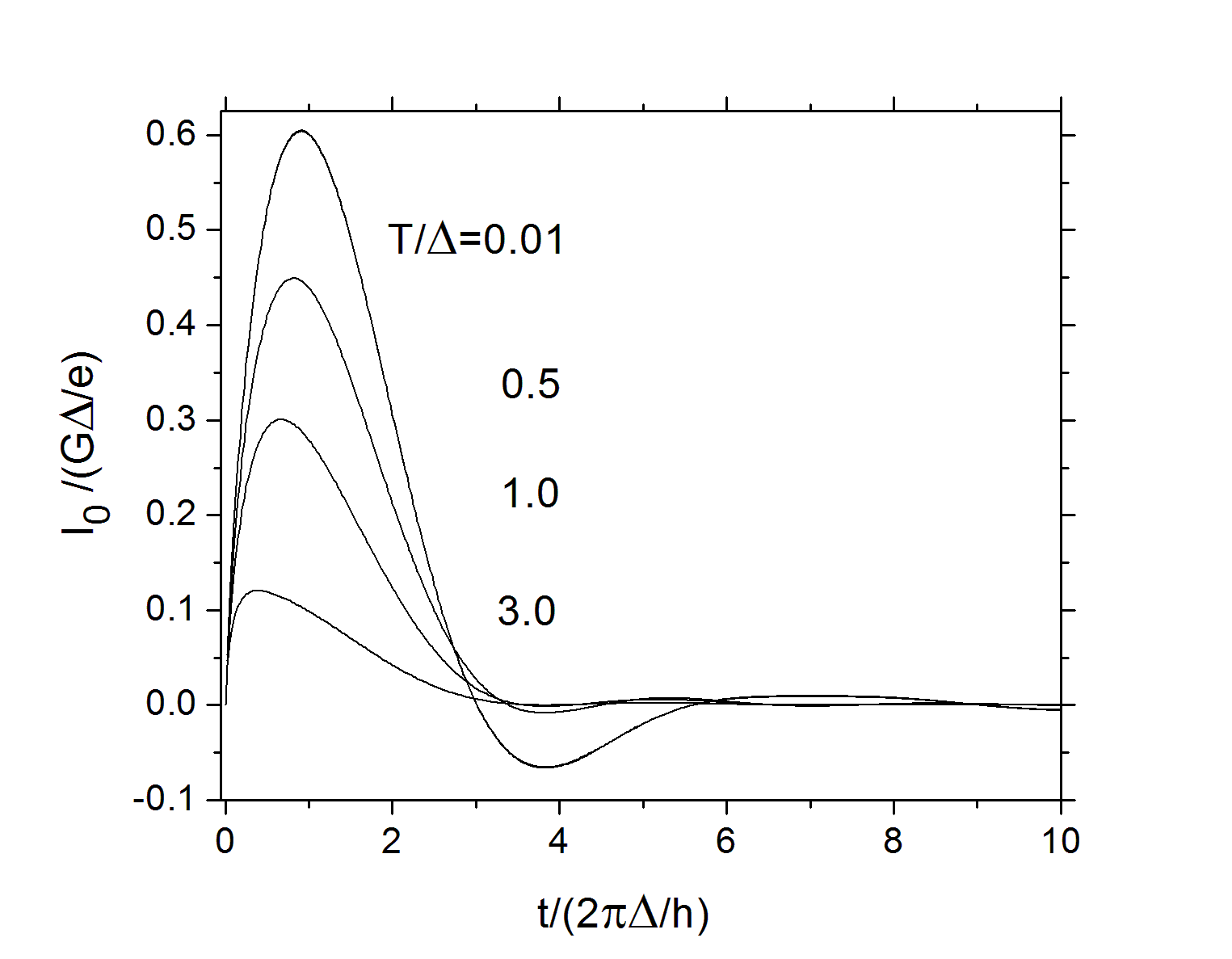}
\caption{{\protect\footnotesize {Oscillating current in a ballistic NS junction driven
by the $\delta$-function pulse of the bias voltage, as a function of time $t$ for several
temperatures $T$. The oscillations are produced by interference of the quasiparticle Andreev-reflection from the gap edges and are suppressed as $1/T$ at large temperatures.} }}
\label{fig2} \end{figure}

Equations (\ref{td}) -- (\ref{TLAR}) are valid for junctions with arbitrary quasiparticle
transparency $D$. They can be simplified further in the two limits of ballistic and the
tunnel junction. For the ballistic junction with $D=1$, the kernel $K(\tau)$ (\ref{corr1})
is given directly by the Fourier transform (\ref{ARF}) of the AR amplitude, and the total
current is:
\begin{eqnarray}
I(t) = G\Big\{V+ \frac{\pi T }{e}\int_{-\infty}^{t} dt'dt'' \frac{\sin[\phi(t')-\phi(t'')]
}{\sinh[\pi T(t'-t'')/\hbar]} \nonumber \\ \cdot \frac{J_1[(t-t')\Delta/\hbar]
J_1[(t-t'')\Delta/\hbar] }{(t-t')(t-t'')} \Big\} \, . \;\;
\label{bal} \end{eqnarray}
In this case, the amplitude $I_0(t)$ (\ref{delta}) can be calculated numerically from
Eq.~(\ref{bal}) and is shown in Fig.~\ref{fig2} for several values of the temperature $T$.
Outside of a small range $t \sim \hbar/T$ near $t=0$, where $I_0$ goes to zero as $t\ln t$, the
lowest curve in Fig.~\ref{fig2} agrees with Eq.~(\ref{TLAR}), which for $D=1$ simplifies to
$I_0(t)=\pi G[\hbar J_1(t\Delta/\hbar)/t]^2/(2eT)$.

\begin{figure}[t]
\centering
\includegraphics[width=0.45\textwidth]{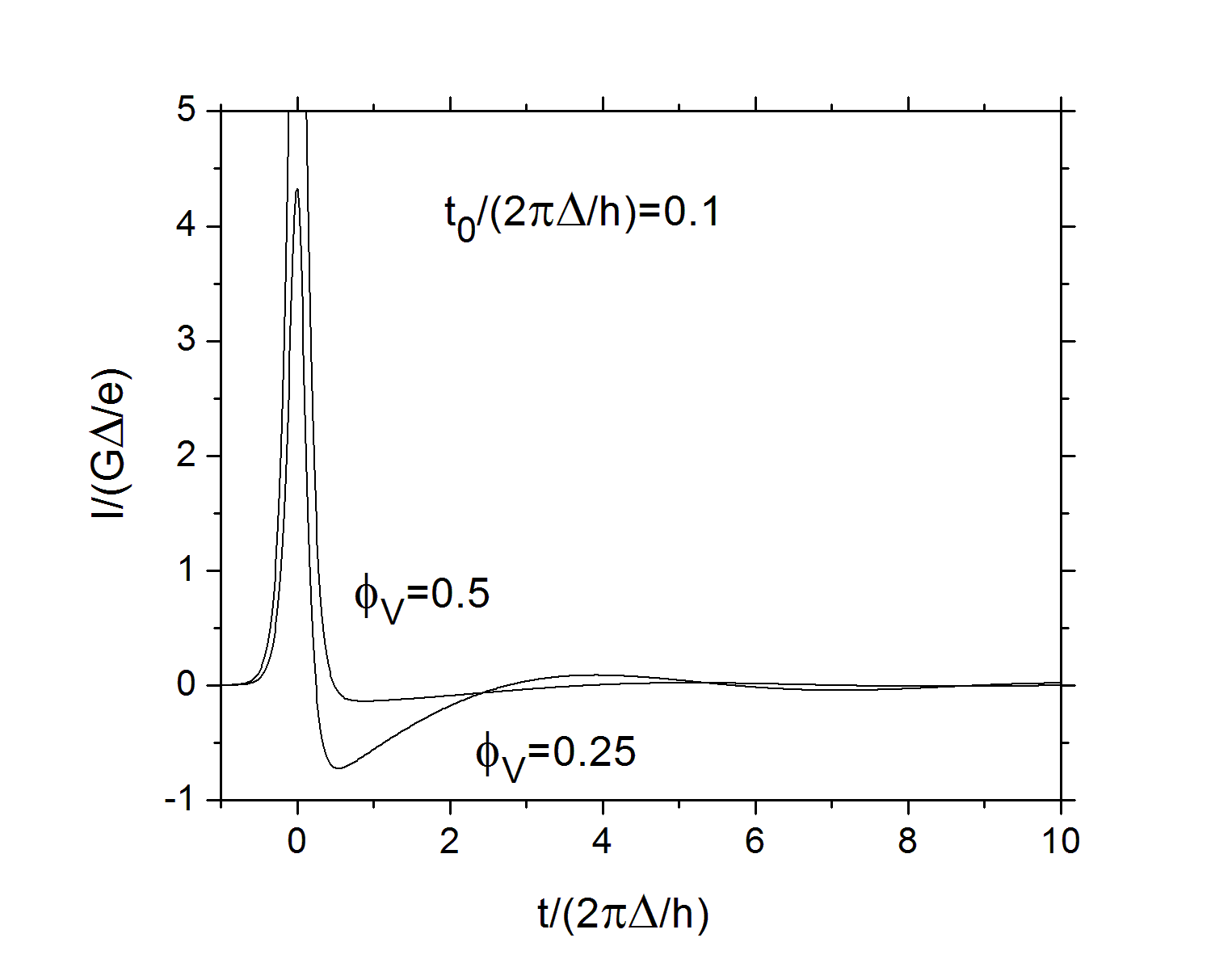}
\caption{{\protect\footnotesize {Current in an NS tunnel junction as a function of time $t$
driven by the two voltage pulses (\ref{pulse}) with two different amplitudes and the same
characteristic time width $t_0$. The curves illustrate the pulse flux control of the
interference current which manifests itself through the oscillating ``tail'' of the current
response to the short voltage pulse. The oscillation amplitude is modulated by the flux
$\phi_V$ carried by the pulse according to Eq.~(\ref{delta}). In the plotted curves, the
oscillations are suppressed for the larger pulse with $\phi_V=0.5$ in comparison to the
smaller pulse with $\phi_V=0.25$. } }}
\label{fig3} \end{figure}

In the tunnel limit $D\ll 1$, one can separate the single-particle contribution $I_T$ to the
current, which is proportional to $D$ and the Andreev-reflection current $I_{AR}$ proportional to
$D^2$. In both of this contributions, one can calculate the kernels $K,L$ taking $D=0$,
i.e. $R=1$ in Eqs.~(\ref{corr3}) and (\ref{corr4}) to get
\begin{equation}
K(\tau)= \frac{1}{2} \Theta (\tau)J_0(\tau) \, ,\;\;\;\; L(\tau)= -\frac{1}{2} \Theta (\tau)
J_1(\tau) \, ,
\label{kltun} \end{equation}
where $J$'s are Bessel functions. Equation ({\ref{TLAR}) for the large-temperature oscillatory
AR current is simplified then accordingly. The single-particle current is described by the last
term in Eq.~(\ref{td}) and explicitly is:
\begin{eqnarray}
I_T(t) = G\Big\{V(t)- \frac{\pi T \Delta}{e\hbar}\int_{-\infty}^{t} dt'
\frac{\sin[\phi(t)-\phi(t')]}{ \sinh[\pi T(t-t')/\hbar]} \nonumber \\
\cdot J_1[(t-t')\Delta/\hbar] \Big\} \, . \;\;
\label{tun} \end{eqnarray}
One can also see directly that expansion in $D$ of the general form of this term does
not have the $D^2$ part, and therefore, the AR part of the current is given directly by
the $D^2$ term in Eq.~(\ref{td}). With the current kernels (\ref{kltun}), the
time-dependent AR current in the tunnel limit is:
\begin{eqnarray}
I_{AR}(t) = \frac{eNTD^2\Delta^2}{4\hbar^3}
\int_{-\infty}^{t} dt'dt'' \frac{\sin[\phi(t')-\phi(t'')]}{ \sinh[\pi T(t'-t'')/\hbar]}
\nonumber \\ \cdot [J_0(\tau-\tau')J_0(\tau-\tau'')-J_1(\tau-\tau')J_1(\tau-\tau'')] . \;\;
\label{ARC} \end{eqnarray}

Finally, an important point to check is how the idealized $\delta$-function limit of the voltage
pulses is approached by the pulses of finite time width. To do this, we adopt the shape of
the bias voltage pulse:
\begin{equation} V(t) =\frac{1}{\pi}\frac{\Phi/t_0}{\cosh t/t_0}\, , \;\;\;
\phi(t) =4 \phi_V \! \arctan e^{t/t_0}\, ,
\label{pulse} \end{equation}
related to the one that can be produced by switching Josephson junctions in the context
of superconductor electronics.  An example of the current in the NS tunnel junction
calculated in the quasiparticle approximation (\ref{tun}) for this pulse shape at low
temperatures is shown in Fig.~\ref{fig3}. This Figure shows that the main qualitative
prediction of the $\delta$-function approximation, flux modulation of the interference
component of the NS current, is reproduced by the pulses of the-not-extremely small
duration $t_0=0.1\hbar/\Delta$.

The work at Stony Brook was supported by the IARPA Supertools program through the Synopsys
and Hypres subcontracts. The authors would like to acknowledge useful discussions with the participants of this program.

\end{document}